\documentclass[a4paper,onecolumn,showpacs,amsmath, amssymb,nofootinbib,aps,floatfix]{revtex4}
\usepackage{amsmath}
\usepackage{amsfonts}
\usepackage{amssymb}
\usepackage[dvips]{graphicx}
\usepackage[dvips]{color}
\usepackage{epsfig}
\input psfig.sty 

\usepackage{bm}                                 
\usepackage{dcolumn}                                
\usepackage{hyperref}

\newcommand{\no}{\mbox{$n_{e_o}$}}

\begin{document}

\title{Complementary cosmological tests of RSII brane models}

\author{R. F. L. Holanda$^{1,3}$\footnote{E-mail: holanda@uepb.edu.br}}

\author{J. W. C. Silva$^4$}

\author{F. Dahia $^{2,3}$\footnote{E-mail: fdahia@df.ufcg.edu.br}}

\address{$^1$Departamento de F\'{\i}sica, Universidade Estadual da Para\'{\i}ba, 58429-500, Campina Grande - PB, Brasil}

\address{$^2$Departamento de F\'{\i}sica, Universidade Federal da Para\'{\i}ba, Jo\~{a}o Pessoa - PB, Brasil}

\address{$^3$Departamento de F\'{\i}sica, Universidade Federal de Campina Grande, 58429-900, Campina Grande - PB, Brasil}

\address{$^4$Departamento de F\'{\i}sica, Universidade Rural do Semi-\'Arido, 59625-900, Paus dos Ferros - RN, Brasil}

\date{\today}

\begin{abstract}

In this paper we explore observational bounds on flat and non-flat
cosmological models in Type II Randall-Sundrum (RSII) branes. In a
first analysis, we consider current measurements of the expansion
rate $H(z)$ (with two priors on the local Hubble parameter) and
288 Type Ia supernovae from the Sloan Digital Sky Survey (within
the framework of the mlcs2k2 light-curve fitting method). We find
that the joint analysis involving these data is an
interesting tool to impose limits on the brane tension density
parameter ($\Omega_{\lambda}$) and that the spatial curvature has
a negligible influence on $\Omega_{\lambda}$ estimates. In order
to obtain stronger bounds for the contribution of the
$\Omega_{\lambda}$ we also add in our analysis the baryon
oscillation peak (BAO) and cosmic microwave background radiation
(CMB) observations by using the so-called CMB/BAO ratio. From this
analysis  we find that the $\Omega_{\lambda}$ contribution is less than
$4.10^{-5}$ (1$\sigma$).

\end{abstract}

\pacs{98.80.-k, 11.25.-w}
\maketitle

\section{Introduction}

Since the formulation of Kaluza-Klein theory in the beginning of
last century, concerning the unification of gravity and
electromagnetism in a five-dimensional space, the study of extra
dimensions have attracted much attention in Physics. Nowadays
there are great interest and much speculation about the existence
of extra dimensions and their influence over our four-dimensional
world, motivated mainly by the braneworld scenarios, in which our
four-dimensional spacetime is viewed as a submanifold
isometrically embedded in a space of higher
dimensions\cite{add1,add2,rs1,rs2,overduim,maartens}.

A relevant aspect of the braneworld scenarios is that matter and
all fields are confined to the brane (below a certain energy level
which is expected to be of Tev order) and only gravity can
propagate into the bulk \cite{add1,add2,rs1,rs2}. Because of this
characteristic, extra dimensions might have large scale as
compared to the tiny Planck scale, assumed by the Kaluza-Klein
theory, without any conflict with the observations so far. In
particular, the RSII model admits the existence of a hidden extra
dimension which has an infinity length \cite{rs2}. If the hidden
extra dimensions are large, in principle, they can be detected
more easily by exploring the signature they might leave in our
four-dimensional space. This could be done by trying to detect
deviations of four-dimensional laws in domains where gravity was
not empirically tested yet. Pursuing this goal, some research
teams have investigated the validity of the inverse square law of
gravity at sub-millimeter scales recently
\cite{newtontest1,newtontest2,newtontest3}. According to them,
there is no evidence of extra dimensions down to a length scale of
about $50\mu m$ \cite{newtontest3}. In cosmology, there are also
many studies about the impact of extra dimensions in the cosmic
evolution considering distinct braneworld
models\cite{maartens,modFried, cosmobrane}. Here we are interested
in discuss some cosmological tests of RSII model.

The RSII model is distinguished from other braneworld models because it is the
first one which, surpassing the paradigm of compact extra dimensions, shows
the possibility that we live embedded in a space with an extra dimension of
infinite size. Despite the fact that gravity can have access to the infinite
extra dimension, a negative cosmological constant $\Lambda_{\left(  5\right)
}$ defined in the bulk ensures the existence of a massless graviton mode,
which is confined to the brane, and the suppression of the massive modes.
These two characteristics together, on their turn, guarantee that the
Newtonian law is recovered at distances much larger than the curvature scale
$\ell=\sqrt{-6/\Lambda_{\left(  5\right)  }}$ established by the
five-dimensional cosmological constant\cite{rs2}.

The laboratory tests of the inverse square law, that have found no
trace of extra dimension, as we already mentioned, then suggest
that the curvature scale $\ell$ could not be greater than
$10^{-4}m$ \cite{newtontest3}. With such a small value, it seems
that the extra dimension will play no significant role in the
recent phase of evolution of the Universe, which, according to
Type Ia supernovae (SNe Ia) data and based on the LCDM model, is
undergoing an accelerated expansion. Needless to say, however,
that any theory should be tested in different scales and by means
of all kind of physical phenomena as much as possible. Hence, in
this paper, we intended to use a recent SNe Ia data
combined with measurements of H(z) as an independent and new test
for the RSII model in the cosmological scale.

As is well known, in the cosmological context, the differences between the
RSII model and the standard cosmological model can be synthesized as
modifications of the Friedmann equation. Admitting that the brane is a
Roberton-Walker three-dimensional space, then the effective field equation of
gravity induced on the brane yields the modified Friedmann equation
\cite{modFried}:%
\begin{equation}
\left(  \frac{H}{H_{0}}\right)  ^{2}=\Omega_{\Lambda}+\Omega_{m}\left(
\frac{\rho}{\rho_{0}}\right)  +\Omega_{\kappa}\left(  \frac{a_{0}}{a}\right)
^{2}+\Omega_{\lambda}\left(  \frac{\rho}{\rho_{0}}\right)  ^{2}+\Omega
_{u}\left(  \frac{a_{0}}{a}\right)  ^{4} \label{modFried}%
\end{equation}
where $a$ is the scale factor, $H$ is the Hubble parameter, $\rho$ is the
density of the dust matter confined to the brane. The number zero as a subscript
indicates the current value of the quantities. The parameters $\Omega
_{\Lambda},\Omega_{m}$ and $\Omega_{\kappa}$ are the density parameters of the
four-dimensional cosmological constant $\Lambda_{\left(  4\right)  }$, of the
matter and of the spatial curvature, respectively. They have the usual
definition $\Omega_{\Lambda}=\Lambda_{\left(  4\right)  }/3H_{0}^{2}$,
$\Omega_{m}=8\pi G\rho_{0}/3H_{0}^{2}$ and $\Omega_{\kappa}=-\kappa/a_{0}%
^{2}H_{0}^{2}$. The new terms are: the density parameter of the
positive brane tension
$\left(  \lambda\right)  $%
\[
\Omega_{\lambda}=\frac{4\pi G\rho_{0}^{2}}{3H_{0}^{2}\lambda}%
\]
and the density parameter of the so-called dark radiation $\Omega_{u}%
=m/a_{0}^{4}H_{0}^{2}$, where $m$ is another free constant of the model.

It is important to emphasize that all terms of Eq. (\ref{modFried}) are
quantities that inhabit the brane, except the dark radiation, which is
essentially a five-dimensional quantity and, therefore, represents a direct
influence of the bulk geometry over the brane \cite{maartens,modFried}. The
physical origin of the dark radiation can be understood by examining this
picture from the perspective of the embedding formalism. It can be shown that
a homogeneous and isotropic three-dimensional space, i.e., a Robertson-Walker
3-space, can be isometrically embedded into the AdS$_{5}$-Shwarzschild
spacetime \cite{Emb1, Emb2}, which is a black hole solution in five dimensions
with a negative cosmological constant. From this point of view, the expansion
of the observable Universe is a consequence of the motion of the brane in the
static AdS$_{5}$-Scwharzschild space. Moreover, it can also be shown that the
trajectory of the brane is dictated by equation (\ref{modFried}), where the
scale factor localizes the brane in the bulk and $m$ is the mass of the bulk
black hole \cite{Emb1, Emb2}.

Other remarkable difference of the equation (\ref{modFried}), in comparison
with the usual Friedmann equation, is that it depends on the energy density
squared. The coefficient of this quadratic term is the parameter density
$\Omega_{\lambda}$ which is inversely proportional to the brane tension. So,
the higher the brane tension the lower the effects of the extra dimension in
our world. This happens because, in the RSII model, the brane tension is
connected to the five dimensional cosmological constant $\Lambda_{\left(
5\right)  }$. A high tension corresponds to a high cosmological constant and,
therefore, a strong suppression of the massive graviton modes. Of course
measurements of the brane tension $\lambda$ by estimates of $\Omega_{\lambda}$ would give us important information
about the influence of extra dimension in the cosmic evolution.

At this point, it is interesting to highlight that $\rho^{2}$ arises too in
cosmological models based on other theories as the Loop Quantum Cosmology
\cite{loop} and non-Riemannian theories \cite{mag}. Therefore, although the RSII brane model is the main motivation of our discussion, the constraints on the density parameter $\Omega_{\lambda}$ we find here could also be used to put bounds on the
parameters of those theories.

The modified Friedmann equation depends on the five density
parameters that should be determined by empirical data. The first confrontation of RSII model with SNe Ia data
was done in Ref. \cite{marek}. The authors found, by using the
Perlmutter samples \cite{perlmutter}, that SNe Ia data provide
weak constraints on the brane tension. The simplest case with
$m=0$ and with only dust matter on the brane yields
$\Omega_{k}=-0.9,$ $\Omega_{\lambda}=0.04,$ $\Omega_{m}=0.59$ and
$\Omega_{\Lambda}=1.27$ as the best fit according to the
$\chi^{2}$ method. Considering these numbers as an unrealistic
result, the prior $\Omega_{m}=0.3$ was added in the flat model. In
this case, $\Omega_{\lambda}<0.037$ at 2$\sigma$ level, by using
the Perlmutter sample A \cite{marek}.

In Ref. \cite{Fay}, another data set of SNe Ia was employed in
order to constrain the parameters of the RSII model. The best fit,
in the flat model with the prior $\Omega_{u}=0\pm0.1$, is
$\Omega_{\lambda}=0.026,$ $\Omega_{m}=0.15$ and
$\Omega_{\Lambda}=0.80,$ by using data of the Riess gold samples
\cite{riess}.
Once again, the constraints were not so strong. For instance, $\Omega_{m}%
\in\lbrack0,0.62]$ at 2$\sigma$ level. The curved model was also investigated,
however, in this case, it was necessary to impose some priors on density
matter ($\Omega_{m}\simeq0.3$) in order to get reasonable values for the
density parameters.

The contribution of the $\rho^{2}$ in cosmology was again
considered in Ref. \cite{marek2}, this time by using a data set of
SNe Ia containing samples from \cite{riess, daly, astier}. In that
paper, the authors also considered other tests like CMB and BAO in
order to constrain the density parameter of the RSII model.
However, taking into account only the SNe Ia data, they found
similar results. Indeed, without imposing any additional prior,
the best fit obtained ($\Omega_{k}=0.34,$ $\Omega_{m}=0.0$ ,
$\Omega_{\lambda}=0.044$ and $\Omega_{\Lambda}=0.646)$ was again
not very reasonable.

Therefore, based on these previous results, it seems that, as SNe
Ia data do not constrain strongly the density parameter of the
matter, we do not find realistic results without assuming some
priors for the matter density. In this paper, we want to show that
this can be avoid if we consider a joint analysis with SNe Ia data
and $H\left( z\right)  $ measurements. We explore flat and
non-flat cosmologies. As we shall see, with only 21 data points,
$H(z)$ test gives supplementary information which naturally put
realistic bounds on the parameters without manipulating priors to
the matter density.

For the sake of simplicity, we are going to admit hereafter that
$m=0$. Of course, this case corresponds to the simplest model, in
which there is no black hole in the original bulk and, therefore,
no dark radiation in the brane. Nevertheless, we have to mention
that there are some papers that have focused their attention on
the analysis of dark radiation in RSII branes also by using SNe Ia
data, see for example\cite{darkrad}.

In order to obtain stronger bounds for the parameters of the RSII model, we
have also considered the CMB and BAO cosmological tests. However, differently from \cite{marek2}, our analysis is
based on the CMB/BAO ratio which is expected to be weakly cosmological model dependent according
to \cite{Sollerman}.

The paper is organized as follows. In section II, we give a short
description of the observational data we have used. The corresponding
constraints on the cosmological parameters are investigated in section III. The
article is ended with a summary of the the main results in the conclusion section.

\section{Data sets}

\subsection{SNe Ia Sample}

In our analyses we have considered the SNe Ia data sample from the Sloan digital sky survey  \cite{Sollerman,Kessler}. This sample is a combination of  103 SNe Ia with redshifts $0.04 < z < 0.42 $, discovered during the first season of the Sloan Digital Sky Survey-II (SDSS-II), with a comprehensive and consistent reanalysis of other datasets \cite{Miknaitis,Astier,riess}, resulting in a combined sample of 288 SNe Ia. These 103 data filled the redshift desert between low- and high-redshift SNe Ia of the previous SNe Ia surveys. Kessler et al. (2009) used two light-curve fitters to obtain  the SNe Ia distance moduli, namely, MLCS2K2 \cite{Jha} and SALT2 \cite{Guy}. In this paper we use the SNe Ia distance moduli obtained via MLCS2K2 calibration since this method does  not have cosmological model dependence \cite{beng}. Furthermore, as is largely employed in literatura, all the results in our analysis (see next section) from SNe Ia data are derived by marginalizing the likelihood function over the nuisance parameter \cite{Rapetti,riess}.

\subsection{H(z) measurements}

In recent years $H(z)$ measurements of the Hubble parameter  have been used to constrain cosmological parameters \cite{simon,zhang,Farooq,Stern,Seikel,Chen}. The idea underlying this approach is based on the measurement of the differential age evolution  of the massive and passively evolving early-type galaxies as a function of redshift, which provides a direct estimate of the Hubble parameter $H(z) = -1/(1 + z )
dz/dt \approx -1/(1 + z)\Delta z/\Delta t$. It is important to stress that direct measurements of $H(z)$ at different redshifts is also possible through measurements of the  radial component of baryonic acoustic oscillations  \cite{Benitez} (for recent $H(z)$ reviews see \cite{zhang}).

In this paper, we have used the 21 $H(z)$ measurements \cite{Stern,Gaz,Moresco} in the redshift range $ 0 < z < 1.75$. The $H_0$  influence on our results is explored by considering two priors in analysis: $68 \pm 2.8$ km/s/Mpc, from a median statistics analysis  of 553 measurements of $H_0$ \cite{Chen}, and $73.8 \pm 2.4$ km/s/Mpc, the most recent one based on HST measurements \cite{Riess2011}. Following \cite{Farooq} we have assumed that the distribution of $H_0$ is a Gaussian with one  standard deviation width  $\sigma_{H_0}$ and mean $\bar{H_0}$, so that we can  build the posterior likelihood function $\mathcal{L}_{H}(\textbf{p})$ that depends only on the parameters $\textbf{p}$ by integrating the product  of exp$(-\chi_H^2 /2)$ and the $H_0$ prior likelihood function
exp$[-(H_0-\bar H_0)^2/(2\sigma^2_{H_0})]$,
\begin{equation}
\label{eq:likely}
\mathcal{L}_{H}(\textbf{p})=\frac{1}{\sqrt{2\pi \sigma^2_{H_0}}}
   \int \limits_0^\infty e^{-\chi_H^2(H_0,\textbf{p})/2}
   e^{-(H_0-\bar H_0)^2/(2\sigma^2_{H_0})} dH_0.
\end{equation}
In this way, to obtain our results from $H(z)$ measurements we maximize the likelihood $\mathcal{L}_H(\textbf{p})$,
or equivalently minimize $\chi_H^2(\textbf{p}) = -2
\mathrm{ln}\mathcal{L}_{H}(\textbf{p})$, with respect to the
parameters $\textbf{p}$ to find the best-fit parameter values to flat and non-flat universes  (see next section).

\subsection{Ratio CMB/BAO}

As it is largelly known, the CMB constrains from the so-called shift parameter $R$ \cite{el} and the BAO measurement $A$ \cite{Eins} have been commonly used to constrain non-standard models, but this approach  is not completely correct since these quantities are derived by using parameters close to standard $w$CDM \cite{Sollerman,el,doran}. In order to avoid some bias in our results we follow Ref.\cite{Sollerman} and use the ratio CMB/BAO. This product cancels out some of the dependence on the sound horizon size at last scattering and a more  model-independent constraint can be achieved in statistical analysis. In our analysis we use two ratio CMB/BAO measurements in redshifts $z=0.2$ and $0.35$, namely \cite{Sollerman} (with one standard deviation error bars)

\begin{eqnarray}   
\frac{d_A(z_*)}{D_V(0.2)}   &=& 17.55\pm0.65,\label{eq:dADV_nodrag}\\
\frac{d_A(z_*)}{D_V(0.35)}  &=& 10.10\pm0.38\\\nonumber
\end{eqnarray}
where $d_{\rm A}(z_*)$ is the comoving angular-diameter distance to  recombination, $z_*$ is the redshift of the last-scattering surface ($z_*=1090$) and  $D_V(z)$ is the so-called  dilation scale.  As emphasized in Ref. \cite{Sollerman} these two poins have a correlation coefficient of $0.337$ \cite{Percival}. In this point, we have to assume that brane models  do not change the physics of the pre-recombination epochs \cite{marek2}.

\begin{figure}
\label{Fig}
\centerline{
\includegraphics[width=2.3truein,height=2.4truein]{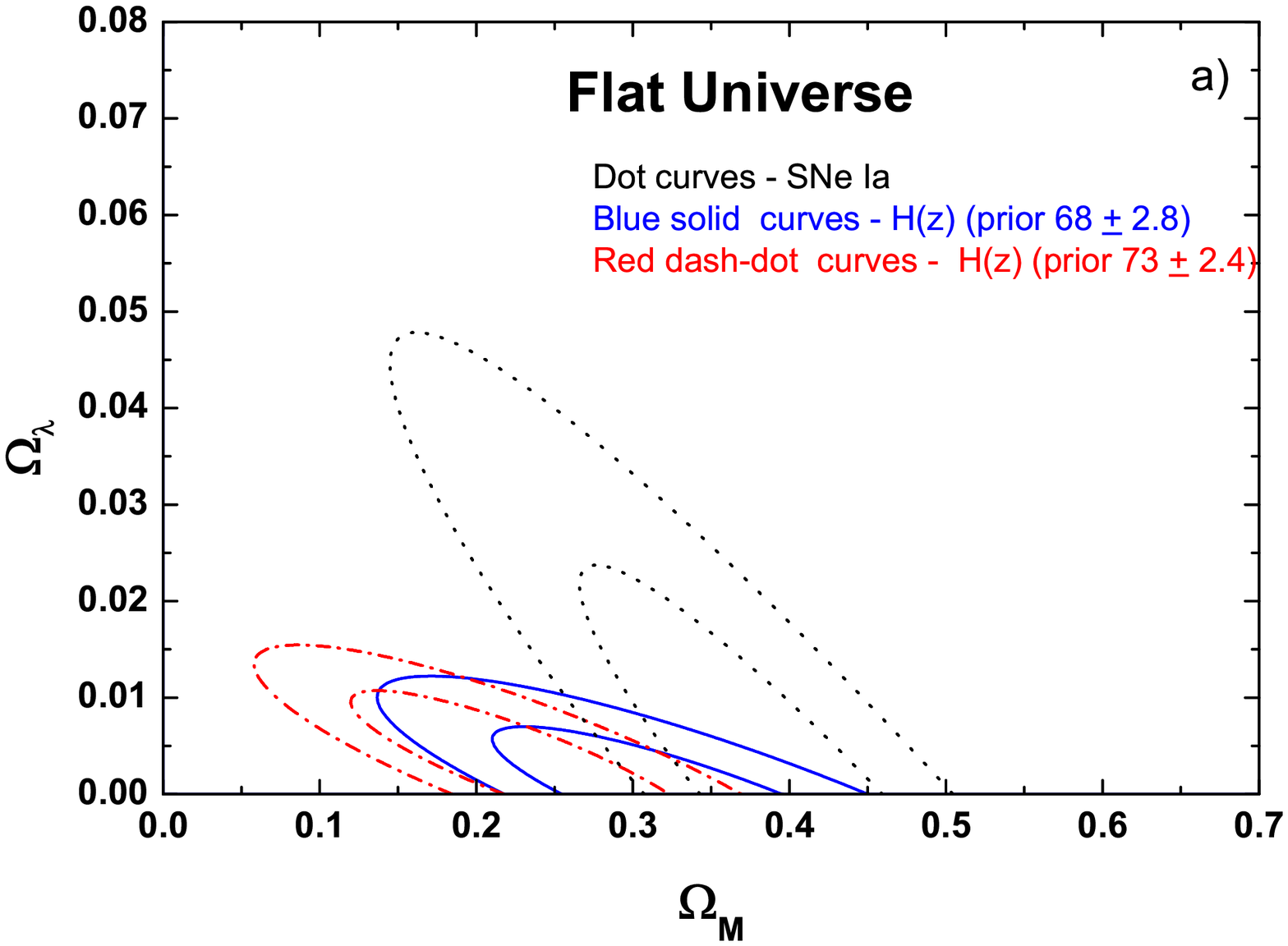}
\includegraphics[width=2.3truein,height=2.4truein]{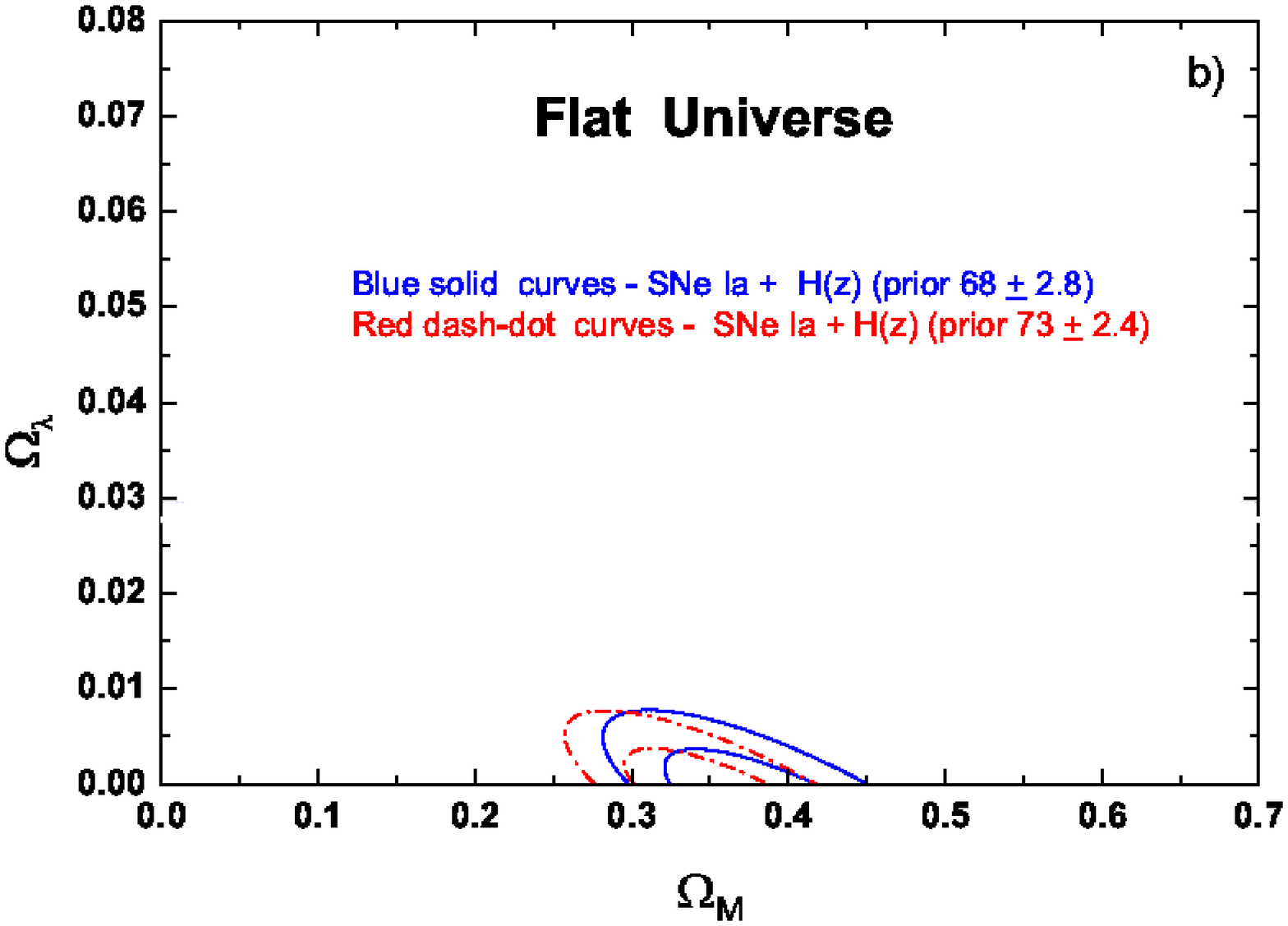}
\includegraphics[width=2.3truein,height=2.4truein]{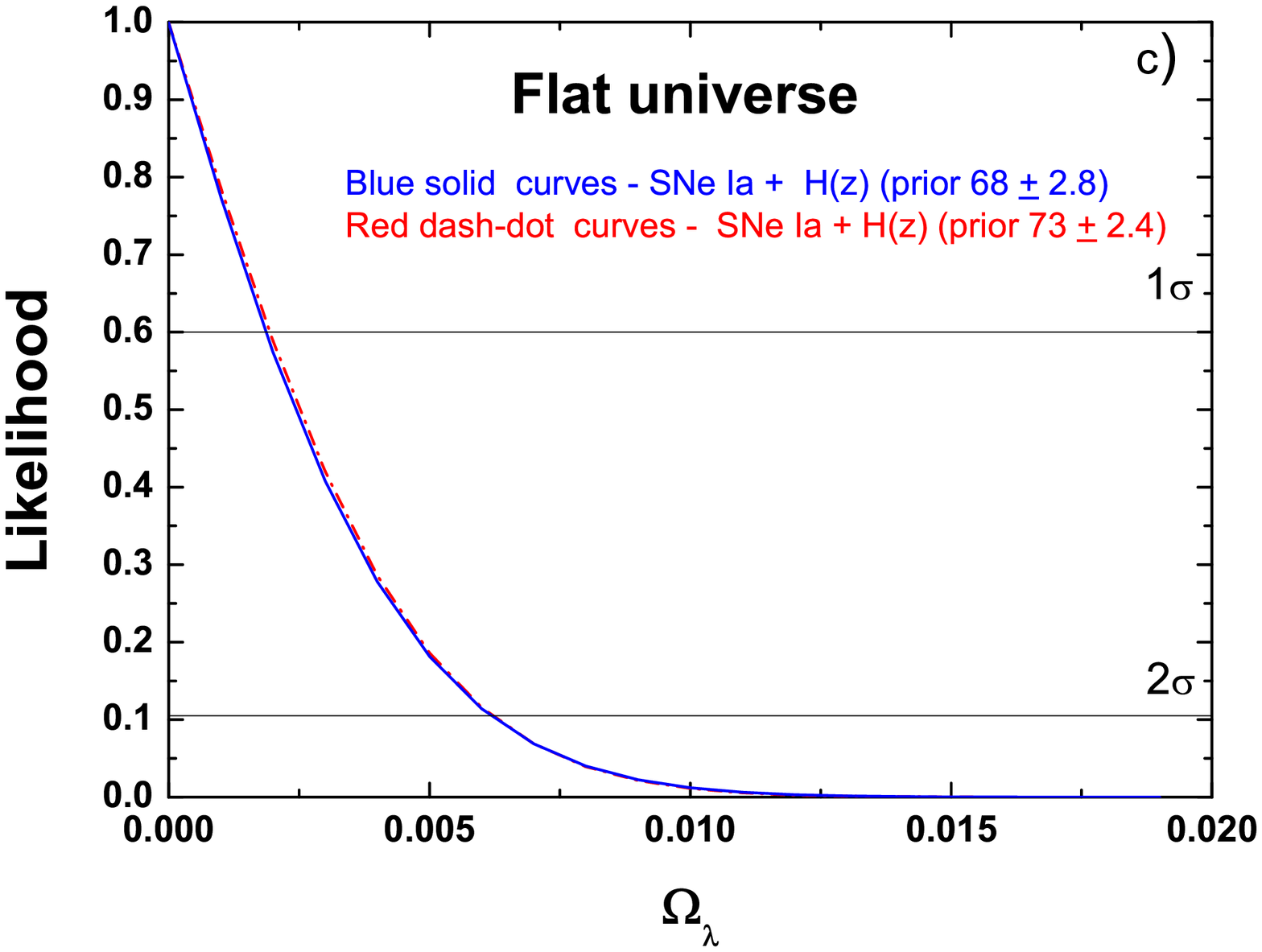}
\hskip 0.1in}

\caption{Confidence contours on the plane $\Omega_{\lambda}-\Omega_{M}$ to flat Universe. Fig. 1a) Contours are drawn to $1\sigma$ and $2\sigma$. The black dots corresponds to bounds from SNe Ia. Through this paper the H(z) priors $68 \pm 2.8$ and $73.8 \pm 2.4$ correspond to blue solid and red dash-dot curves, respectively. Fig. 1b) shows the contours  to $1\sigma$ and $2\sigma$ from our joint analysis. Fig. 1c)  displays the likelihoods to $\Omega_{\lambda}$.}
\end{figure}

\section{Results}

\subsection{Flat Universe}

The results of our statistical analyses to flat universe are shown in Figures 1a, 1b and 1c. Figure 1a shows contours of $1\sigma$ and $2\sigma$  on the $\Omega_{\lambda} -\Omega_{M}$ plane when SDSS (MLCS2K2) and H(z) data points are considered separately (through this paper the priors $68 \pm 2.8$ and $73.8 \pm 2.4$ correspond to blue solid and red dash-dot curves, respectively). In this case $\Omega_{\Lambda}$ can be find by using simply $\Omega_{\Lambda}=1-\Omega_{\lambda} -\Omega_{M}$. By considering two free parameters, we can find from the SNe Ia sample ($1\sigma$): $\Omega_M = 0.37 \pm 0.11$ and $\Omega_{\lambda} = 0.0 + 0.022$ ($\chi^2 = 237.84$). On the other hand, from H(z) data points we find: $\Omega_{M} = 0.32 \pm 0.1$ and $\Omega_{\lambda} = 0.0 + 0.006$ ($\chi^2 = 15.87$), and $\Omega_{M} = 0.24 \pm 0.12$ and $\Omega_{\lambda} = 0.0 + 0.01$ ($\chi^2 = 16.9$) when the priors $68 \pm 2.8$ km/s/Mpc and $73.8 \pm 2.4$ km/s/Mpc are used on $H_0$, respectively. Note that no significant conflict between the results from $H(z)$ priors is found and the bounds on $\Omega_{\lambda}$  from 21 H(z) data points are tighter  than ones from 288 SNe Ia. Moreover, the contours from SNe Ia and H(z) analyses  are almost ortogonal each other and a joint analysis can be used to impose tighter limits on the space parameters.

Figure 1b shows our results from a joint analysis. We find $\Omega_{M} = 0.37 \pm 0.05$ and $\Omega_{\lambda} = 0.0 + 0.0033$ ($\chi^2 = 255.45$) and $\Omega_{M} = 0.33 \pm 0.04$ and $\Omega_{\lambda}=0.0 + 0.0031$ ($\chi^2 = 260.45$) by using SNe Ia + H(z) with the priors $68 \pm 2.8$ km/s/Mpc and $73.8 \pm 2.4$  km/s/Mpc on $H_0$, respectively.

Panel 1c display the likelihood for the $\Omega_{\lambda}$ parameter, marginalizing on $\Omega_{M}$. In this case we obtain $\Omega_{\lambda}=0.0 + 0.0018$ to both priors.

\subsection{Non-Flat Universe}

In this section, we  allow deviations from flatness  to  prove the robustness of the previous $\Omega_{\lambda}$ constraints using SNe Ia and H(z) data points. The results of our statistical analyses  are shown in figures 2a, 2b and 2c (here we marginalize over $\Omega_{\Lambda}$). Figure 2a show contours of $1\sigma$ and $2\sigma$  on the $\Omega_{\lambda} -\Omega_{M}$ plane when SDSS (MLCS2K2) and H(z) data points are considered separately. In this case we can not constrain  $\Omega_{M}$  and $\Omega_{\lambda}$ simultaneously and, for instance: $\Omega_{M}=0.50$ and $\Omega_{\lambda}=0.005$, and $\Omega_{M}=0.20$ and $\Omega_{\lambda}=0.035$ are permitted with high degree of confidence by using the SNe Ia sample. Furthermore, $\Omega_{\lambda}=0$ is excluded at $2\sigma$ to $\Omega_{M}=0.1$ and it is permitted at $1\sigma$ to $\Omega_{M}=0.3$. Similar conclusion can be done to results from H(z).  According with these comments a joint analysis is necessary in order to break the degenerescency on the ($\Omega_{\lambda}$,$\Omega_{M}$) plane.

Figure 2b displays the results by using a joint analysis  involving SNe Ia and H(z) data points. We find ($1\sigma$):  $\Omega_{M} = 0.38 \pm 0.09$,  $\Omega_{\lambda}=0.0 + 0.0012$ ($\chi^2 = 255.33$)  and $\Omega_{M} = 0.41 \pm 0.08$, $\Omega_{\lambda} = 0.0 + 0.001$  ($\chi^2 = 260.21$) with the priors $68 \pm 2.8$ km/s/Mpc and $73.8 \pm 2.4$ km/s/Mpc on $H_0$, respectively. On the other hand, by marginalizing on $\Omega_M$ we find: $\Omega_{\Lambda} = 0.65 \pm 0.17$  and $\Omega_{\Lambda} = 0.80 \pm 0.18$, respectively.

Panel 1c display the likelihood for the $\Omega_{\lambda}$ parameter, marginalizing on $\Omega_{M}$ and $\Omega_{\Lambda}$. In this case we obtain ($1\sigma$) $\Omega_{\lambda}=0.0 + 0.0018$ and $\Omega_{\lambda}=0.0 + 0.0016$  to $68 \pm 2.8$ km/s/Mpc and $73.8 \pm 2.4$ km/s/Mpc $H_0$ priors, respectively. Therefore, the limits on $\Omega_{\lambda}$ are equivalent to the flat case and we see that the geometry has a negligible influence on  estimates for this combination of data.

Finally, in order to obtain tighter limits on space parameters  we add in our analyses two points of the so-called ratio CMB/BAO. In figures 3a, 3b and 3c we show our results. We find (at $1\sigma$):  $\Omega_{M} = 0.28 \pm 0.03$, $\Omega_{\lambda}=0.0 + 0.0001$  ($\chi^2 = 262.88$) and $\Omega_{M} = 0.28 \pm 0.03$,  $\Omega_{\lambda} = 0.0 + 0.00006$  ($\chi^2 = 264.21$) with the priors $68 \pm 2.8$ and $73.8 \pm 2.4$, respectively. By marginalizing on $\Omega_M$ we find: $\Omega_{\Lambda} = 0.65 \pm 0.10$ and $\Omega_{\Lambda} = 0.68 \pm 0.10$ for each case, respectively.  It is very important to stress that the limits derived by using SNe Ia + H(z) + CMB/BAO are more cosmological model independent than previous analyses where  the $A$  and $R$  quantities were used \cite{marek2}. The likelihood curves are plotted in figure 3c. We obtain $\Omega_{\lambda}=0.0 + 0.00004$ and $\Omega_{\lambda}=0.0 + 0.00002$ ($1\sigma$) to $68 \pm 2.8$ and $73.8 \pm 2.4$ priors, respectively.

\begin{figure}
\label{Fig}
\centerline{
\includegraphics[width=2.3truein,height=2.4truein]{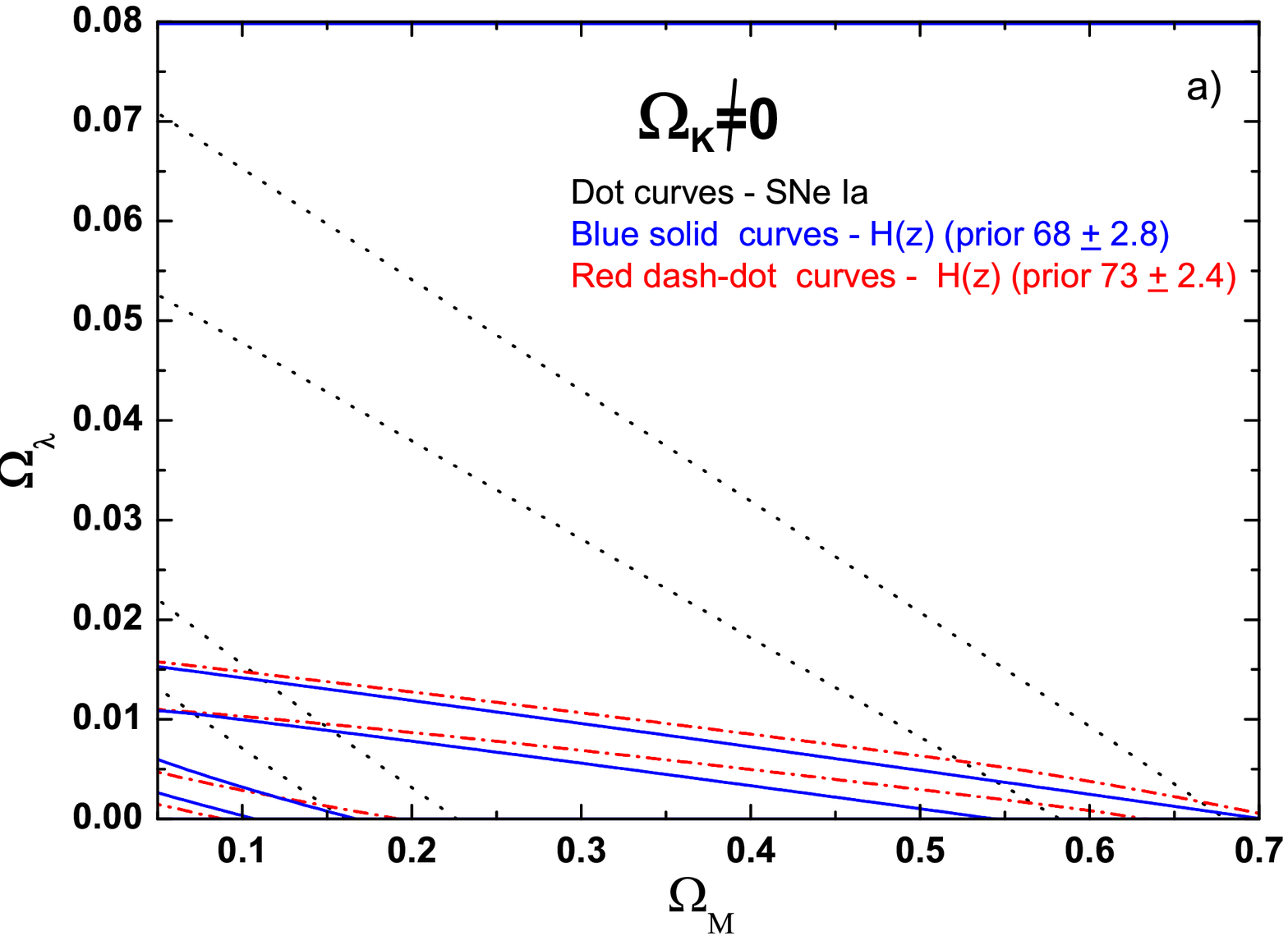}
\includegraphics[width=2.3truein,height=2.4truein]{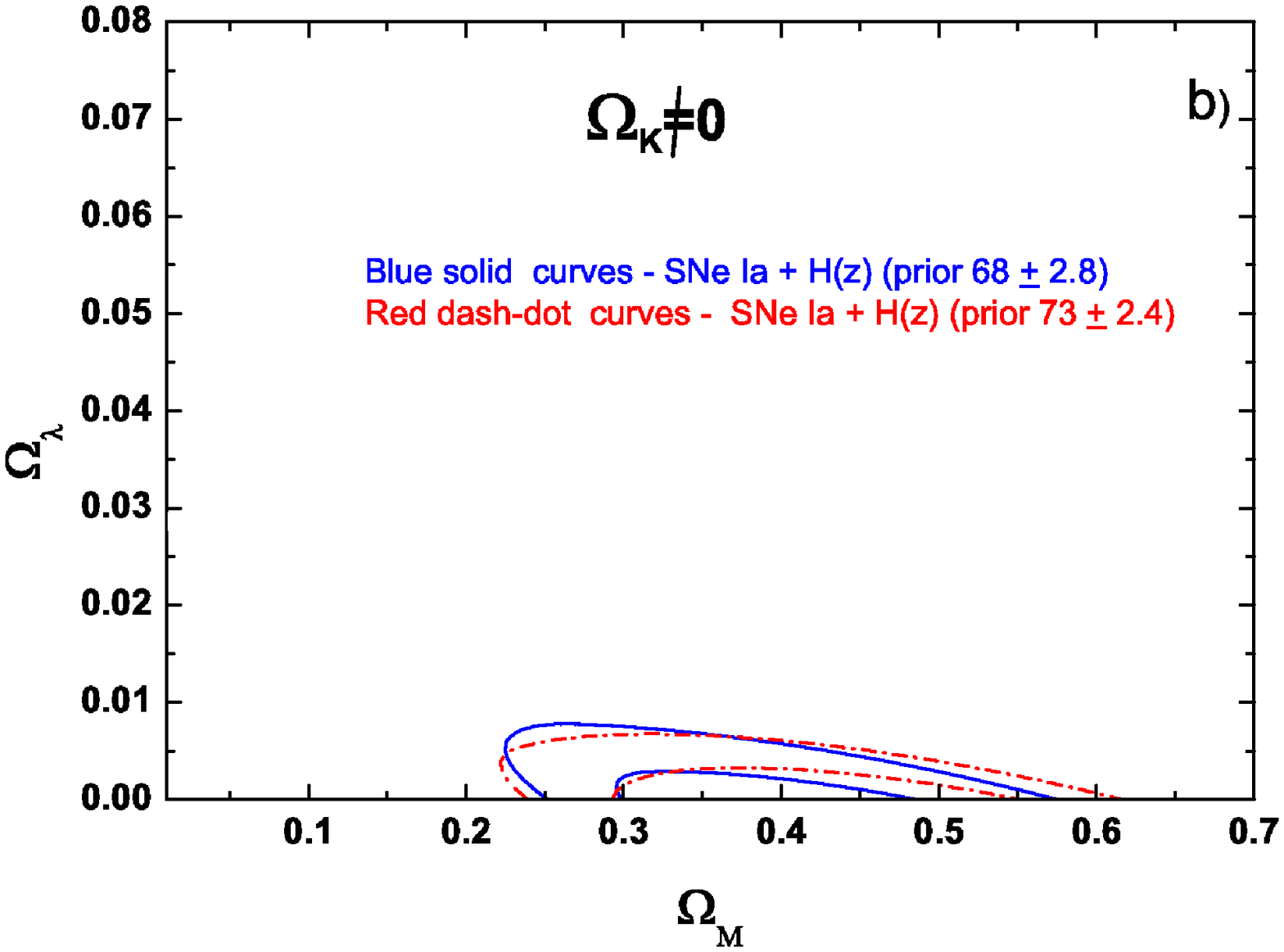}
\includegraphics[width=2.3truein,height=2.4truein]{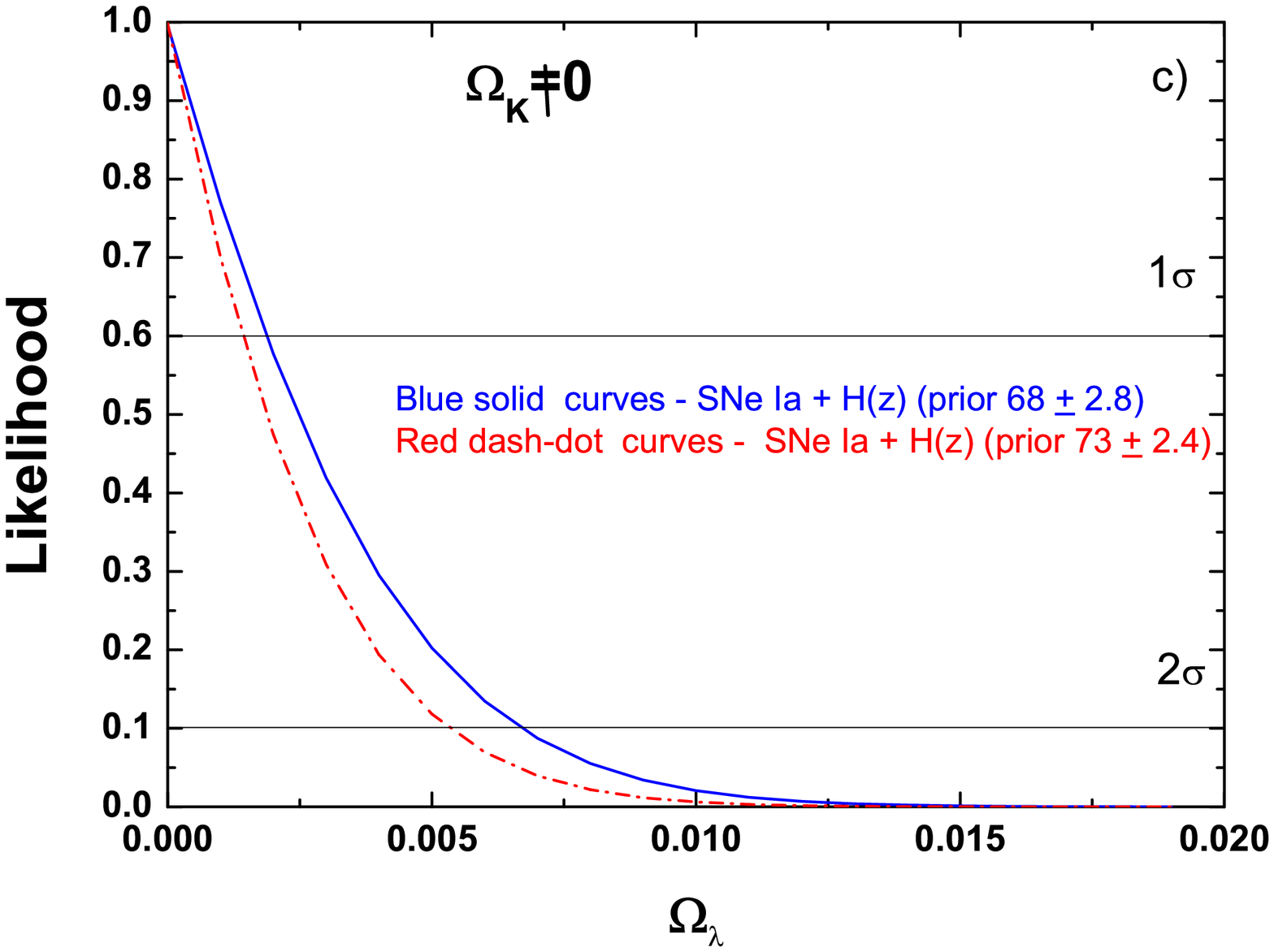}
\hskip 0.1in}
\caption{Confidence contours on the plane $\Omega_{\lambda}-\Omega_{M}$ to flat universe. Fig. 2a) Contours are drawn to $1\sigma$ and $2\sigma$. The black dots corresponds to bounds from SNe Ia. Through this paper the H(z) priors $68 \pm 2.8$ and $73.8 \pm 2.4$ correspond to blue solid and red dash-dot curves, respectively. Fig. 2b) shows the contours  to $1\sigma$ and $2\sigma$ from our joint analysis. Fig. 2c)  displays the likelihoods to $\Omega_{\lambda}$.}
\end{figure}

\section{Conclusions}

\label{sec:conclusions} In this paper, we have studied
observational bounds on cosmologies based on the flat and non-flat
RSII brane models. Previous papers showed that SNe Ia data do not
provide strong constraints on the parameters of RSII model.
Moreover, the best fits did not give realistic values unless some priors to the matter and curvature densities were assumed.
However, as we have shown here, when we consider the joint
analysis of 288 SNe Ia (SDSS by using MLCS2K2 method) and 21 H(z)
measurements we obtain consistent results with other observations
\cite{Hinshaw} for matter density without imposing such priors.
Moreover, the constraints on $\Omega_{\lambda}$ are weakly
dependent on the curvature parameter and $H_{0}$ priors ($68 \pm
2.8$ and $73.8 \pm 2.4$ km/s/Mpc), we have obtained at $1\sigma$:
$\Omega_{\lambda} \approx 0.0 + 0.002$  to flat and non-flat
Universes (see Figs. 1c and 2c). It is important to
emphasize that in the near future it will be possible to determine
around 1000 $H(z)$ values of the Hubble parameter at a 15\%
accuracy level \cite{simon,alcaniz}. This fact shows that upcoming
$H(z)$ measurements plus SNe Ia may become a useful tool to
impose constraints on $\Omega_{\lambda}$.

We also have added the so-called CMB/BAO ratio in our analysis in
order to obtain tighter limits on $\Omega_{\lambda}$. As is known,
this ratio is more model independent than analyses done by using
the $A$ and $R$ quantities which have parameters close to standard
$w$CDM in their derivations. The joint analysis (SNe
Ia+H(z)+CMB/BAO) gives the following values at $1\sigma$:
$\Omega_{\lambda}=0.0 + 0.00004$ and $\Omega = 0.0 + 0.00002$ with
the priors $68 \pm 2.8$ and $73.8 \pm 2.4$, respectively. { In
terms of the brane tension $(\lambda),$ which is a fundamental
parameter of the model, these bounds impose that $\lambda>[0.01$
$eV]^{4}$.}

{ At this point it is interesting to compare the present results with other
tests. Since in the RS braneworld models the modified Friedman equation
involves a $\mathcal{O}\left(  \frac{\rho}{\lambda}\right)  $ correction term
 the bounds on the tension found here are much weaker than the lower limit
put by the analysis of BBN $\lambda>[1$ $MeV]^{4}$
\cite{maartens,bbn}, for instance. The reason is very simple, the
present energy density of universe is extremely small and the
contribution of such term becomes more important at early times,
and hence observations concerning BBN can put better constraints
on $\lambda$. Following this reasoning, one might expect that the
analysis of the formation of large scale structure can constitute
another a way to obtain independent bounds for the brane tension.
The equations that describes the evolution of the cosmological
perturbations in the brane are very complex in the RSII model
\cite{maartens,p1,p2,p3,p4,p5,p6,p7,p8,p9,p10,p11,p12,p13}. In
comparison with the GR equations, there are
two major differences: the presence of $\mathcal{O}\left(  \frac{\rho}%
{\lambda}\right)  $ correction terms and the coupling between the
perturbations in the brane and the fluctuations of the
five-dimensional bulk geometry (the so-called KK-modes). As the
brane is embedded in the bulk, the fluctuations of the bulk
geometry influence the brane geometry back. This means that the
KK-modes act as another source for the perturbation in the brane
\cite{maartens,p1,p2,p3,p4,p5,p6,p7,p8,p9,p10,p11,p12,p13,p14}.
Therefore, perturbations in the brane cannot be studied separately
from the 5D-perturbations in the bulk, except in some special
cases \cite{p1,p13,p15,p16}. In order to deal with such a complex
differential system usually one resorts to numerical analysis.
Some results show that scalar perturbations (as the density
perturbation) with a wavenumber lower than a critical value
$k_{c}$ are amplified in comparison with the GR values, in the
radiation era \cite{p17}. The critical value that is defined as
$k_{c}=H_{c}a_{c}$ corresponds to the critical mode that enters
the horizon at the critical epoch which is established by the
condition $H_{c}\ell=1$. As $k_{c}$ depends on $\ell$, which is
related to the brane tension
($\lambda=\frac{3\pi}{4}\frac{c^{4}}{G}\frac{1}{\ell^{2}}$), in
principle, we can use data set about the large scale structure to
constrain the brane tension.

Finally, we have to mentioned that the most stronger bound obtained so far
comes from laboratory tests, which put the following lower limit to the brane
tension: $\lambda>\left[  10\text{ }TeV\right]  ^{4}$\cite{maartens,
newtontest1,newtontest2,newtontest3}. In comparison with these results, we see
that ours are much weaker. However, we should have in mind that our analysis
is based on different physical phenomena and on physical processes that
happened at a different cosmological epoch as compared to BBN and to those
processes involved in the laboratory tests. Thus, we may say that the method
proposed here constitute an independent test to check the consistency of the
model at large scale. Moreover, in the near future, as more and larger data
sets with smaller statistical and systematic uncertainties become available,
the present method will put tighter limits on the value of the brane tension.}

\begin{figure}
\label{Fig}
\centerline{
\includegraphics[width=2.3truein,height=2.4truein]{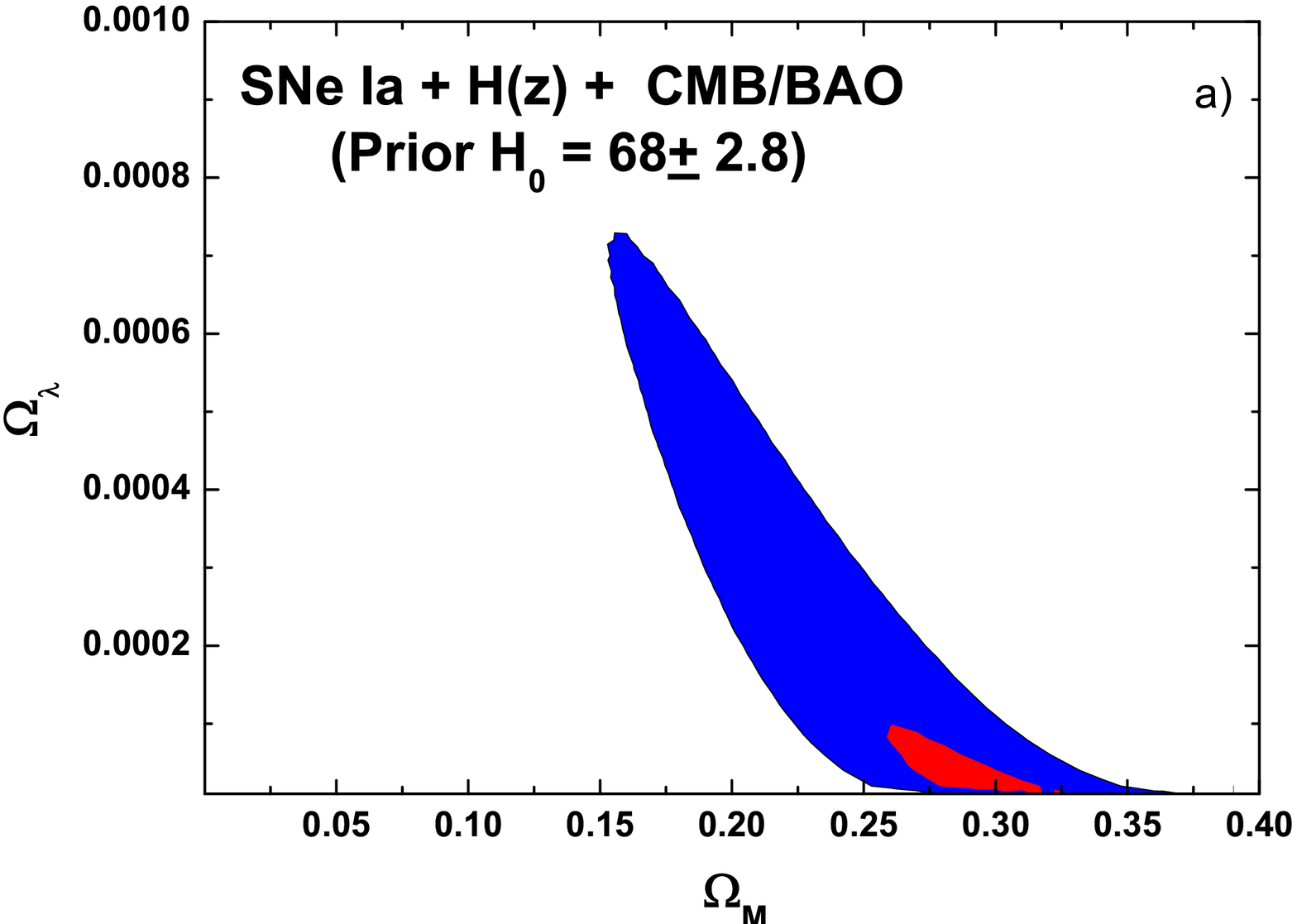}
\includegraphics[width=2.3truein,height=2.4truein]{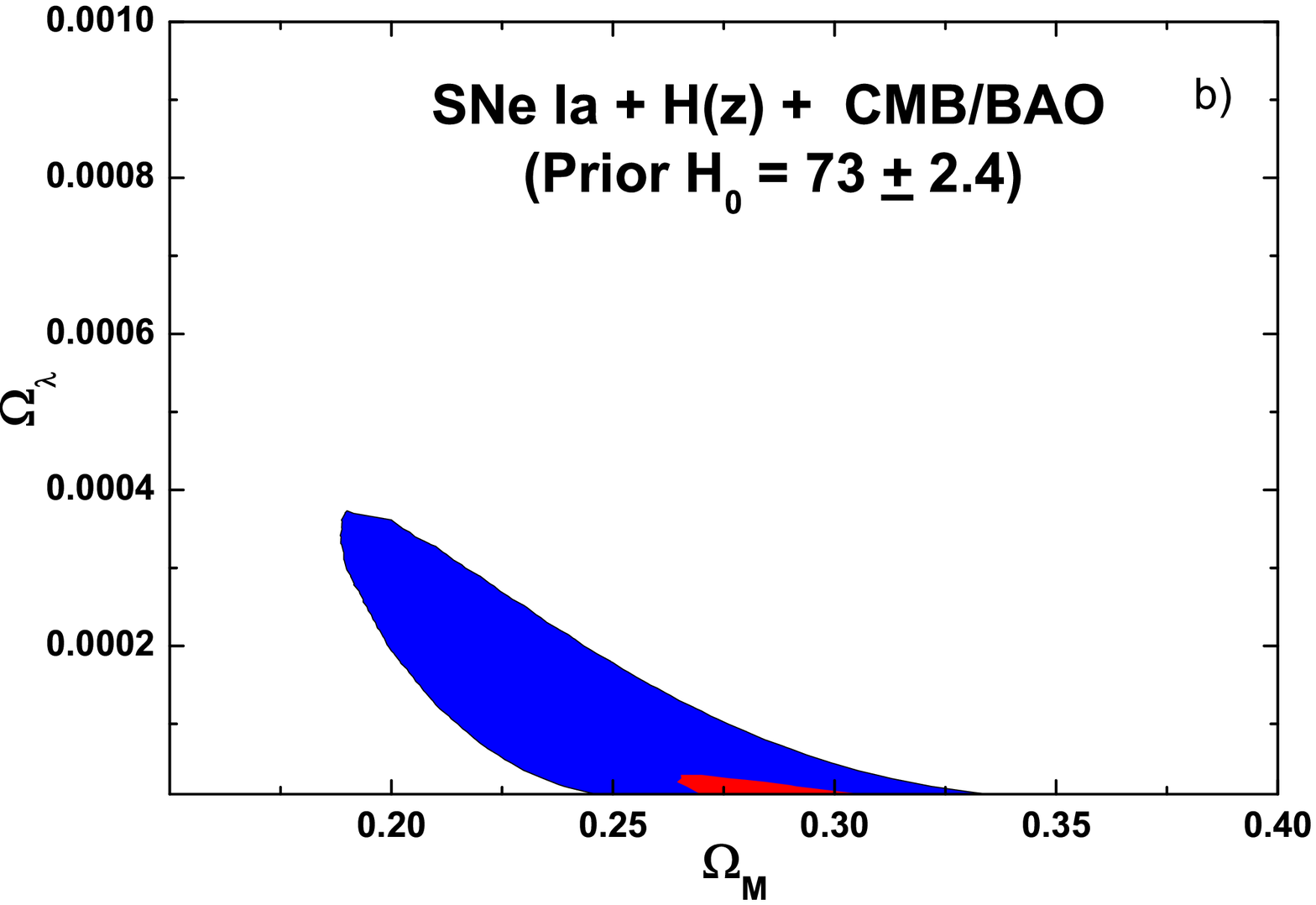}
\includegraphics[width=2.3truein,height=2.4truein]{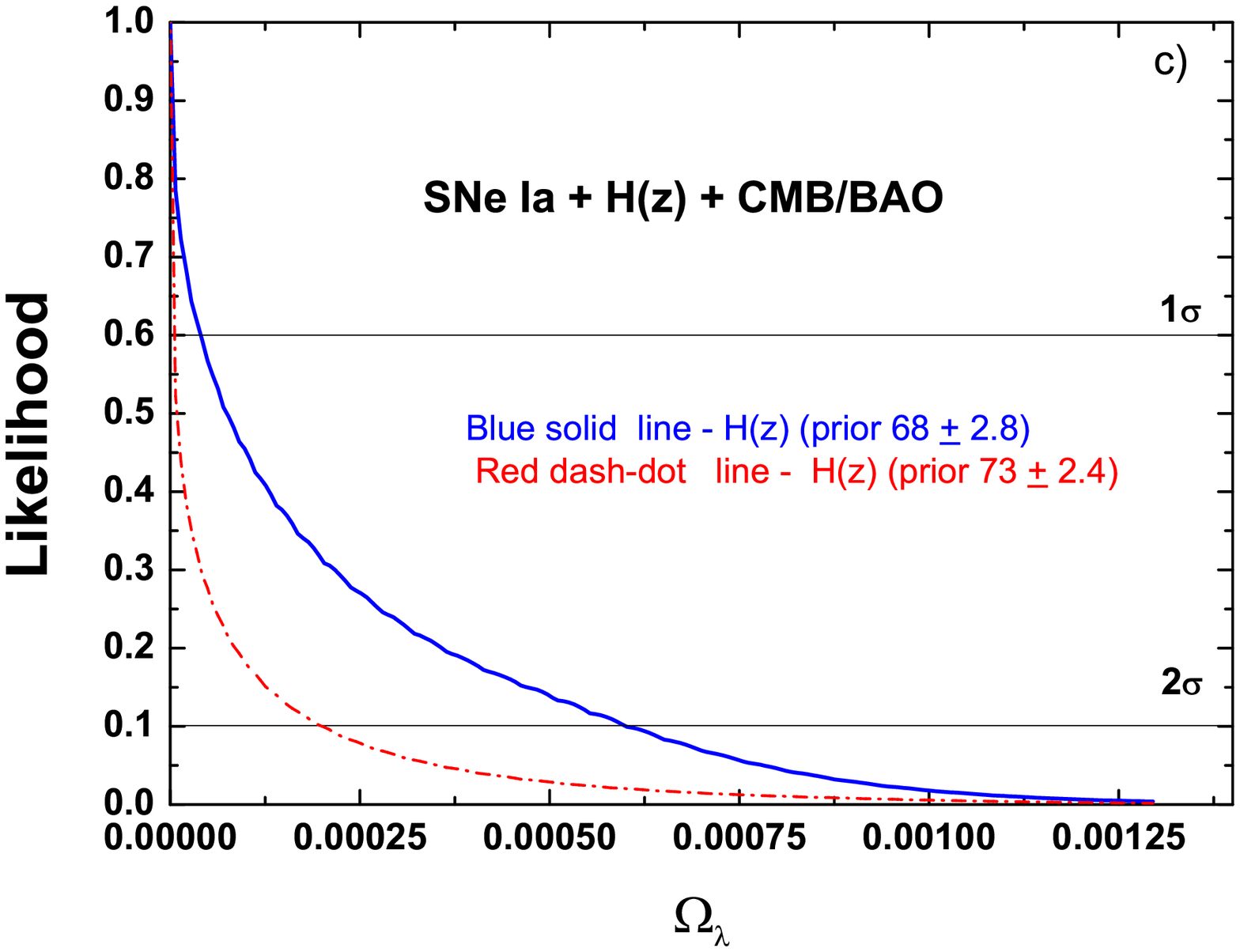}
\hskip 0.1in}
\caption{Confidence contours on the plane $\Omega_{\lambda}-\Omega_{M}$ to non-flat universe by using SNe Ia + $H(z)$ + BAO/CMB ratio. Fig. 2a) Contours are drawn to $1\sigma$ and $2\sigma$, $H_0$ prior corresponds to $68 \pm 2.8$ km/s/Mpc. Fig. 2b) Contours are drawn to $1\sigma$ and $2\sigma$,  $H_0$ prior corresponds to $73.8 \pm 2.4$  km/s/Mpc. Fig. 3c) displays the likelihoods to $\Omega_{\lambda}$.}
\end{figure}

\section{Acknowledgments}

R. F. L. Holanda thanks INCT-A for the grants under which this
work was carried out.

\end{document}